\documentclass[aps,pra,twocolumn,amsmath,amssymb,footinbib,showpacs,longbibliography,superscriptaddress]{revtex4-2}

\usepackage[english]{babel}
\usepackage{latexsym}
\usepackage{graphics}
\usepackage{subfigure}
\usepackage{epsfig}
\usepackage{xcolor}
\usepackage{hyperref}
\usepackage{braket} 
\usepackage[T1]{fontenc}
\usepackage[latin9]{inputenc}
\usepackage{amstext}
\usepackage{amssymb}
\usepackage{graphicx}
\usepackage{esint}
\usepackage{braket}
\usepackage{babel}
\usepackage{amsmath}
\usepackage{siunitx}
\usepackage{autobreak}
\usepackage{comment}
\usepackage{booktabs}
\usepackage{tabularx}
\usepackage{placeins}

\hypersetup{
	colorlinks=true,
	citecolor=blue,
	linkcolor=red,
	urlcolor=black
}


\newcommand{\e}{\textrm{e}}
\newcommand{\ie}{i.e.}

\newcommand{\D}{\mathrm{d}}
\newcommand{\pa}{\partial}

\newcommand{\inftyint}{\int_{-\infty}^{\infty}}
\newcommand{\calP}{\mathcal{P}}

\newcommand{\calK}{\mathcal{K}}

\newcommand{\aHR}{a}

\newcommand{\kB}{k_{\mathrm{B}}}
\newcommand{\kF}{k_\textrm{\tiny F}}

\newcommand{\EF}{E_\textrm{\tiny F}}
\newcommand{\TF}{T_\textrm{\tiny F}}
\newcommand{\kmax}{k_\textrm{max}}

\begin{document}

\title{
Thermodynamics and Tomonaga-Luttinger liquid behavior of the quantum 1D hard rod model
}

\author{Shengjie Yu}
\affiliation{CPHT, CNRS, Ecole Polytechnique, IP Paris, F-91128 Palaiseau, France}

\author{Zhaoxuan Zhu}
\affiliation{Beijing National Laboratory
for Condensed Matter Physics, Institute of Physics, Chinese Academy of Sciences, Beijing 100190,
China}

\author{Laurent Sanchez-Palencia}
\affiliation{CPHT, CNRS, Ecole Polytechnique, IP Paris, F-91128 Palaiseau, France}

\date{\today}

\begin{abstract}
The one-dimensional hard rod model describes impenetrable bosons with finite diameter, extending the Lieb-Liniger model to systems with excluded volume interactions.
Here, we investigate the thermodynamics of quantum HRs using Yang-Yang theory, path integral quantum Monte-Carlo calculations, and Luttinger liquid theory.
We first discuss the behavior of characteristic thermodynamic quantities, exhibiting deviations to the Lieb-Liniger model for sufficiently high densities,
with excellent agreement between analytical and numerical results.
We then show that the hard rod model exhibits Tomonaga-Luttinger liquid behavior across a wide range of parameters, at zero and finite temperature, as unveiled by correlation functions.
The Tomonaga-Luttinger parameter and thermal length can be extracted by fitting correlation functions to Tomonaga-Luttinger liquid theory, hence demonstrating a robust method for thermometry.
This work provides a comprehensive study of strongly correlated hard rod systems at finite temperatures,
with applications to quantum wires, spin chains, and ultracold atoms.
\end{abstract}

\maketitle

\section{Introduction}\label{sec:introduction}
Hard-sphere models describe hard-core particles with a finite diameter.
They form a central class of classical statistical models for modeling universal thermodynamic properties of repulsive particle gases, 
with applications to liquid-gas transitions and fluid dynamics~\cite{hansen2006,rosenbluth1954,dyre2016,mulero2008, percus1976equilibrium, olla2022diffusive},
as well as generalized hydrodynamics~\cite{castro2016emergent,bertini2016transport,doyon2017drude,biagetti2024three,hubner2026diffusive}.
Classical models, as well as their quantum counterparts, have been extensively studied with direct applications to
rigid molecules and colloidal systems~\cite{royall2024},
strongly repulsive particles in gaseous helium on carbon nanotubes~\cite{pearce2005,calbi2001},
spin-ion compounds~\cite{michelsen2019},
polarized hydrogenoids~\cite{vidal2016},
and ultracold atoms~\cite{rigol2004emergence,mazzanti2008ground,astrakharchik2005beyond,GDeRosi2017thermodynamic,GDeRosi2019Beyond,GDeRosi2024Analytic}.
Notably, hard spheres are also reminiscent of the excluded volume associated with Rydberg blockade in atomic systems~\cite{naik2024}.
In one dimension (1D), the hard rod (HR) model is a natural extension of the Lieb-Liniger model, which describes 1D bosons with contact interactions~\cite{lieb1963a,lieb1963b}.

In quantum systems,
the reduced dimensionality exaggerates quantum correlations, leading to unique phenomena in 1D,
such as the
absence of Bose-Einstein condensation,
fermionization of interacting Bose gases, 
and the breakdown of quasiparticle excitations~\cite{giamarchi2004,cazalilla2011}.
Quantum systems in 1D are also highly susceptible to perturbations with characteristic lengths commensurate with the average inter-particle distance,
leading to the celebrated pinning transition in arbitrary weak periodic potentials~\cite{haldane1980,haldane1981,giamarchi1997}
and Bose glass transitions in disordered or quasiperiodic systems~\cite{giamarchi1987,giamarchi1988,vidal2001,roux2008,roscilde2008,yao2019,yao2020}.
Unlike in 3D, quantum correlations in 1D bosonic systems increase with decreasing particle density, making them an ideal testbed for studying strongly correlated regimes in dilute systems.
This applies to a variety of physical systems, including
electronic quantum wires, spin chains,
organic conductors,
and tightly confined ultracold atoms~\cite{giamarchi2004,cazalilla2011}.
Recent experimental observations have confirmed a number of phenomena, such as Tonks-Girardeau (TG) physics~\cite{kinoshita2004,paredes2004}, absence of thermalization in integrable systems~\cite{kinoshita2006}, spin-charge separation~\cite{kim1996,kim2006}, Mott transitions in shallow periodic potentials~\cite{haller2010,boeris2016}, disorder-induced enhancement of coherence~\cite{yao2024}, and the observation of the Tan contact~\cite{yao2018,huang2024direct}.

Another asset of quantum 1D systems is that they are particularly suited to exact or quasi-exact theoretical approaches.
Most of their properties are captured by the universal harmonic fluid theory, known as Tomonaga-Luttinger liquids (TLL)~\cite{tomonaga1950,luttinger1963,haldane1981b}.
The latter predicts algebraically decaying correlation functions with exponents determined by the sole TLL parameter, which is however model dependent.
Moreover, quantum 1D systems are amenable to quasi-exact numerical approaches based on  tensor-network representations~\cite{white1993,schollwock2005,schollwock2011}.
Last but not least, 1D offers a variety of exactly-solvable strongly-correlated quantum models,
including the celebrated Lieb-Liniger (LL) model~\cite{lieb1963a,lieb1963b} and a number of descendents~\cite{sutherland2004}.

The HR model has been previously studied in several works. The standard approach for the ground state involves mapping onto the LL model with infinite repulsive interactions (TG gas) by excluding the volume occupied by the HRs and shifting the ordered coordinates of the particles accordingly~\cite{nagamiya1940statistical}. 
Beyond the ground state, more advanced approaches are required.
The Bethe ansatz (BA) and thermodynamic Bethe ansatz (TBA) solutions have been derived in Refs~\cite{sutherland1971groundstate,sutherland1971quantum,wadati2002one,vsamaj2013introduction}. 
Moreover, the dynamical structure factor and TLL behavior of HRs have studied at zero temperature using variational Monte Carlo~\cite{mazzanti2008ground,motta2016dynamical}.

In this work, we extend these studies to finite temperatures.
We first briefly review the BA and TBA solutions for the HRs.
Then, we perform quantum Monte Carlo (QMC) calculations at finite temperatures, finding excellent agreement with the exact TBA solutions.
We also compute one- and two- body correlation functions and find results in excellent agreement with TLL behavior.
The unknown coefficients from the TLL theory are then extracted from fits to the QMC results.
We show that using these fits, the universal properties of TLL allow for a thermometry of 1D systems described by HRs.
In addition, we discuss the behavior of the static structure factor, which displays non-analytical peaks at zero temperature and power-law scaling of the peaks as a function of temperature.
Finally, possible applications to Rydberg atom systems are outlined.

\section{Bethe ansatz solution}
\label{sec: exact solution}
We first briefly review the BA solution for the ground state (GS) and excitations.
Similar derivations were first presented in Refs~\cite{sutherland1971groundstate,sutherland1971quantum,wadati2002one,vsamaj2013introduction}, see also Ref.~\cite{motta2016dynamical}.
The Hamiltonian for 1D bosons with pairwise interactions is given by
\begin{equation}\label{eq: hamiltonian}
\hat{H} = -\frac{\hbar^2}{2m}\sum_{j=1}^N \frac{\pa^2}{\pa x_j^2} + \sum_{\ell > j} V(x_\ell - x_j),
\end{equation}
where $m$ is the particle mass, $N$ is the total number of particles, and $x_j$ denotes the center position of particle $j$.
Equation~(\ref{eq: hamiltonian}) describes a family of models determined by the scattering potential $V(x)$.
For HRs, it reads as $V(x)=\infty$ for $|x| \leq \aHR$ and $V(x)=0$ for $|x| > \aHR$,
which describes  impenetrable particles with finite diameter $\aHR$.

The BA wavefunction is written as
\begin{equation}\label{eq: BA wavefunction}
\Psi(\mathbf{x}) =
\sum_{\calP\in S_N} A_\calP \exp\bigg( i\sum_{j=1}^N k_{\calP_j}x_j \bigg),
\end{equation}
for \( | x_{j+1} - x_j | \geq a \), and \( \Psi(\mathbf{x}) = 0 \) otherwise.
The latter guarantees the HR impenetrability condition, which forces the distance between consecutive particles to exceed \( \aHR \).
Here, \( S_N \) is the symmetric group over \( \{1, \ldots, N\} \), and \( \mathbf{x} = (x_1, \ldots, x_N) \) represents the ordered particle coordinates.
Since  \( \Psi(\mathbf{x})\) vanishes for \( x_{j+1} - x_j < a \), the energy is fully kinetic and reads as
$E=\sum_j \hbar^2 k_j^2/2m$.
To determine the coefficients \( A_{\mathcal{P}} \) and 
quasi-momenta \( k_j \), the continuity condition
requires the wavefunction to vanish at the HR boundaries,
\ie~$\Psi(x_1,...,x_j,x_j+\aHR,...,x_N)=0$, for all $ j \in \{1,...,N\}$.
It yields the relations
$A_{(j,j+1)\calP} = -e^{i\Theta(k_{\calP_{j+1}} - k_{\calP_j})} {A_\calP}$,
for any $j$,
where \( \Theta(k) = k a \) is the phase shift of two-body scattering, defined in the same way as the Lieb-Liniger model.
These equations are solved by
\begin{equation}
A_\calP = (-1)^\calP A \exp\Big[ -i\sum_j k_{\calP_j}(j-1)\aHR \Big].
\end{equation}
Periodic boundary conditions (PBCs) then yield the BA equation
\begin{equation}\label{eq: BA eqn}
k_j L = 2\pi I_j + \sum_{\ell \neq j} \Theta(k_j - k_\ell)
\end{equation}
for any $j \in \{1,...,N\}$,
where 
the quantum numbers \( I_j \) take integer or half-integer values for odd or even \( N \), respectively.
The parity of $N$ contributes since the PBCs involve (N-1) interchanging of momenta, leading to a phase of $(N-1)\pi$. 
The set of $N$ quantum numbers $\{I_j\}$ characterizes the eigenstate uniquely.
For HRs, the linear form of \( \Theta(k) \) allows for an explicit solution. By summing all Eqs.~(\ref{eq: BA eqn}) for all $j$'s, one finds the total momentum
$Q=\sum_j k_j=2\pi I /L$, with $I= \sum_j I_j$.
Then, inserting this formula for $Q$ and $\Theta(k)=ka$ into Eq.~(\ref{eq: BA eqn}),
one finds
\begin{equation}\label{eq: quasi-momenta}
k_j = \frac{2\pi I_j - 2\pi I \aHR/L }{L-N\aHR},\, I= \sum_j I_j.
\end{equation}
Note that the dimensionless quantities \( k_j \aHR \) and \( E \aHR^2 m / \hbar^2 \) are universal functions of \( \aHR / L \) and \( N \). The TG gas, which describes infinitely repulsive, point-like particles, is found in the limit $\aHR \rightarrow 0$, and we recover the known solution \( k_j^{\textrm{\tiny TG}} = 2 \pi I_j / L \)~\cite{girardeau1960}.

\subsection{Ground state}
\label{sec:BA.GS}
The ground state is obtained by minimizing the energy.
The quadratic form of the latter implies that the set of quasi-momenta \( k_j \) should be symmetric, and the total momentum \( Q = \sum_j k_j = 2 \pi I / L \) should vanish, \ie~\( Q = 0 \) and \( I = 0 \).
The quantum numbers \( I_j \) take all integer or half-integer values in the set \( \{-\frac{N-1}{2}, \ldots, \frac{N-1}{2}\} \), hence creating a Fermi-like sea with maximum momentum $\kmax=\pi n/(1-n\aHR)$, where $n=N/L$ is the particle density,
in the thermodynamic limit.
The energy per particle is then
\begin{equation} \label{eq: energy2}
\frac{E_0}{N} \approx \frac{\hbar^2 \pi^2 n^2}{6m(1-na)^2}.
\end{equation}
Compared to the TG gas, the quasi-momenta are rescaled as $k_j=k_j^{\textrm{\tiny TG}}/(1-na)$ and the energy as $E_0 = E_0^{\textrm{\tiny TG}}/(1-na)^2$. The energy divergence at $na=1$ is due to the complete impenetrability of HRs, which implies the condition $N\aHR \leq L$, \ie~$na \leq 1$ in the thermodynamic limit.
Note, however, that the zero-temperature chemical potential, $\mu=\pa E_0/\pa N$, is rescaled differently, due to the dependence of the energy rescaling on the total particle number \( N \).
One finds
\begin{equation} \label{eq: mu}
\mu = \frac{\hbar^2\pi^2n^2}{2m}\times\frac{1-n\aHR / 3}{(1-n\aHR)^3},
\end{equation}
\ie~$\mu = \mu^{\textrm{\tiny TG}} \times {(1-n\aHR / 3)}/{(1-n\aHR)^3}$ with $\mu^{\textrm{\tiny TG}}=\hbar^2\pi^2n^2/2m$, the chemical potential of the TG gas.

\subsection{Excitation spectrum}
\label{sec:BA.Excitations}
As usual in BA, elementary excitations are generated
by changing one of the \( I_j \)'s, creating two distinct branches~\cite{lieb1963a,lieb1963b}.
Particle-type excitations (p, upper branch) are created by promoting a particle from the highest populated momentum \( k_N = \kmax \) to a larger value \( q > \kmax \)
while hole-type excitations (h, lower branch) involve moving a particle from within the range of populated momenta to just above the largest quasi-momentum.
In both cases, although a single quantum number $I_j$ is changed,
all quasi-momenta are shifted since $I$ turns from $0$ to $\neq 0$, see Eq.~(\ref{eq: quasi-momenta}).
Although, the shift is infinitesimal in the thermodynamic limit, it plays a crucial role because it affects all the momenta and has a macroscopic impact~\cite{motta2016dynamical}.

Adding a particle or a hole implies adding or removing a quasi-momentum $\pm q_0$ and shifting the other quasi-momenta so that $k' \rightarrow k'+f_\mathrm{p,h}(k')$.
The total momentum for a particle or hole excitation then reads as
\begin{equation}\label{eq:excitation_p}
q = \pm q_0 + \int_{-\kmax}^{\kmax} J_\mathrm{p,h}(k') \D k',
\end{equation}
where the plus and minus signs correspond to the particle and hole types respectively, and the energy as
\begin{equation}\label{eq:excitation_e(p)}
\varepsilon_\mathrm{p,h}(q) = \pm \left(\frac{\hbar^2 q_0^2}{2m} - \mu\right) + \frac{\hbar^2}{2m} \int_{-\kmax}^{\kmax} 2kJ_\mathrm{p,h}(k')\D k',
\end{equation}
where the chemical potential $\mu$ compensates the energy of adding or removing a particle
and \( J_\mathrm{p,h}(k) = \rho(k) f_\mathrm{p,h}(k) L \) is the density of quasi-momentum shift, where $\rho(k)$ is the density of quasi-momenta in the ground state.
The latter is determined using Eq.~(\ref{eq: BA eqn}), which yields the self-consistent equation
\begin{equation} \label{eq:excitation_J(k)}
2\pi J_\mathrm{p,h}(k) = -\int_{-\kmax}^{\kmax} \calK(k-k') J_\mathrm{p,h}(k') \D k' \mp [\pi + \Theta(q_0-k)],
\end{equation}
with $\calK(k) = \D \Theta/\D k$.
For HRs, the constant kernel, \( \mathcal{K}(k) = a \), allows again for an analytical solution,
\begin{equation} \label{eq:J(k)_HR}
J_\mathrm{p,h}(k) = \mp \frac{(1-n\aHR)(q_0\aHR + \pi) -k\aHR}{2\pi}.
\end{equation}
Inserting this result in Eqs.~(\ref{eq:excitation_p}) and (\ref{eq:excitation_e(p)}), we find the analytic expressions
\begin{equation}
q = \pm (1-n\aHR)\left(q_0 - \kmax\right) , \label{eq:p_HR}
\end{equation}
and
\begin{equation}  \label{eq:e(p)_HR}
\varepsilon_\mathrm{p,h} (q) = \frac{\hbar^2 q(2\pi n \pm q)}{2m (1-n\aHR)^2},
\end{equation}
for the momentum and energy, respectively. We hence recover the results in Refs~\cite{vsamaj2013introduction,motta2016dynamical}.

\begin{figure}[tbp!]
\centering
\includegraphics[width=\columnwidth]{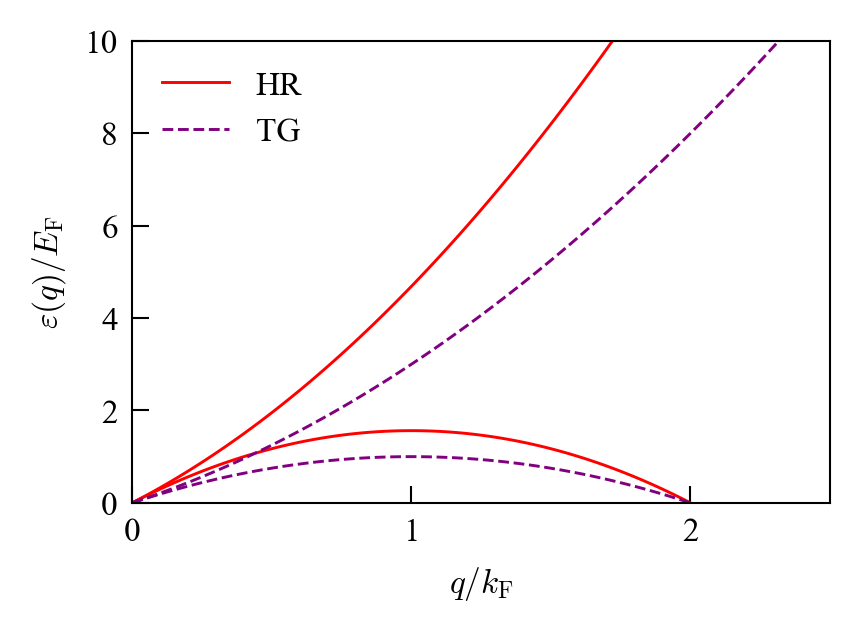}
\caption{
Dispersion relation of elementary particle (upper branch) and hole (lower branch) excitations for the HR (solid red line) and the TG (dashed purple line) models at $n\aHR=0.2$. Here $\EF=\pi^2 \hbar^2 n^2/2m$ is the Fermi energy.}
\label{fig: excitations}
\end{figure}
The dispersion relations of the elementary excitations for the HR, and TG models at $na=0.2$ are shown in Fig.~\ref{fig: excitations}.
Both models exhibit qualitatively similar behaviours but significant quantitative differences.
The TG model is recovered from Eq.~(\ref{eq:e(p)_HR}) for $\aHR \rightarrow 0$, and, for HRs, the same energy rescaling $\varepsilon_\mathrm{p,h} (q) = \varepsilon_\mathrm{p,h}^\mathrm{TG}(q)/(1-n\aHR)^2$ as for the ground state.
In particular, the hole branch has a momentum cut-off at the Fermi momentum $\kF = \pi n$, for both HR and TG models.

\section{Thermodynamics}
\label{sec: thermodynamic}
We now turn to the thermodynamics of the HR model at finite temperature.
The different rescalings of the elementary excitation energies and chemical potential with $n\aHR$ implies that no such simple rescaling can be applied here and a complete TBA approach must be derived~\cite{sutherland1971groundstate,sutherland1971quantum,wadati2002one,vsamaj2013introduction}.
This will allow to determine the thermodynamics of the model and benchmark QMC calculations.

To write the thermodynamic Bethe ansatz (TBA) of the HR model, we work along the lines of Yang and Yang~\cite{yang1969thermodynamics}.
In brief, we define $ \rho(k) $ as the density of filled quasi-momenta and $ \rho_h(k) $ the density of holes, subject
to the normalization condition
\begin{equation} \label{eq:constraint_rho_rhoh}
\rho(k) +\rho_h(k) = \frac{1}{2\pi} - \frac{1}{2\pi} \inftyint \calK(k-k') \rho(k') \D k',
\end{equation}
where $\calK(k)\equiv \D \Theta/\D k=a$. Minimizing the free energy written in terms of the dressed energy,
\begin{equation} \label{eq:constraint_rho_rhoh_epsilon}
\epsilon(k) \equiv k_B T \ln [\rho_h(k) / \rho(k)],
\end{equation}
one finds the self-consistent Yang-Yang equation~\cite{wadati2002one,vsamaj2013introduction}
\begin{equation}
\epsilon(k)  = -\mu + \frac{\hbar^2 k^2}{2m} + \kB T \int \frac{\D k'}{2\pi} \calK(k-k') \ln \left[ 1 + e^{-\frac{\epsilon(k')}{\kB T}} \right].
\label{eq:yang-yang}
\end{equation}
This equation is solved numerically for $\epsilon(k)$ using an iterative method.
The latter is significantly simplified for HRs compared to the Lieb-Liniger model by the fact that $\mathcal{K}$ is a constant, which lifts the $k$-dependence of the integral in Eq.~(\ref{eq:yang-yang}), see Appendix~\ref{appendix:dressed-energy} for details.
Then, using the numerical result for $\epsilon(k)$ in Eqs.~(\ref{eq:constraint_rho_rhoh}) and (\ref{eq:constraint_rho_rhoh_epsilon}),
we find $\rho(k)$ and $\rho_h(k)$, as well as the free energy
\begin{equation}
F=N\mu - L\int \frac{\D k}{2\pi} \ln \left[ 1 + e^{-\epsilon(k) / \kB T} \right].
\end{equation}
The particle density $ n = \int_{-\infty}^{\infty} \mathrm{d}k \rho(k) $ is found from Eq.~(\ref{eq:constraint_rho_rhoh}),
which reduces to $ \rho(k) + \rho_h(k) = (1 - n \aHR) / 2 \pi $, leading to
\begin{align}
n\aHR = & \frac{\chi_2(\mu,T)}{1 + \chi_2(\mu,T)},
\label{eq: density}
\end{align}
with $\chi_2(\mu,T) = \aHR \inftyint \frac{\D k}{2\pi} \frac{1}{1 + \e^{\epsilon(k)/\kB T}}$.
This formula may be used to compute the compressibility using the standard thermodynamic relation,
\begin{equation}
\kappa
\equiv \left( \frac{\pa n}{\pa \mu} \right)_{T}
= \frac{(1-n\aHR)^3 \chi_3(\mu,T)}{\kB T \aHR},
\label{eq: compressibility}
\end{equation}
with
$\chi_3(\mu,T) = \aHR \inftyint \frac{\D k}{2\pi} \frac{ e^{\epsilon(k)/\kB T} }{ \left(1 + e^{\epsilon(k)/\kB T}\right)^2 }$.
Both Eqs~(\ref{eq: density}) and (\ref{eq: compressibility}) are explicitly derived in Appendix~\ref{appendix:density-compressibility}.
They provide characteristic quantities of the HR gas, namely the equation of state and the compressibility, respectively.
Other thermodynamic quantities, not discussed hereafter, are similarly found from derivatives of the free energy.

Then we turn to QMC simulations, as these provide us with more information than BA does, such as correlation functions.
First of all, we benchmark the QMC against the exact TBA results.
Our implementation works in the grand canonical ensemble with fixed temperature and chemical potential, and worm-algorithm updates~\cite{boninsegni2006}.
The specific propagator for the HR interaction is given in Appendix~\ref{appendix:propagator}.
\begin{figure}[tbp]
\centering
\includegraphics[width=1\columnwidth]{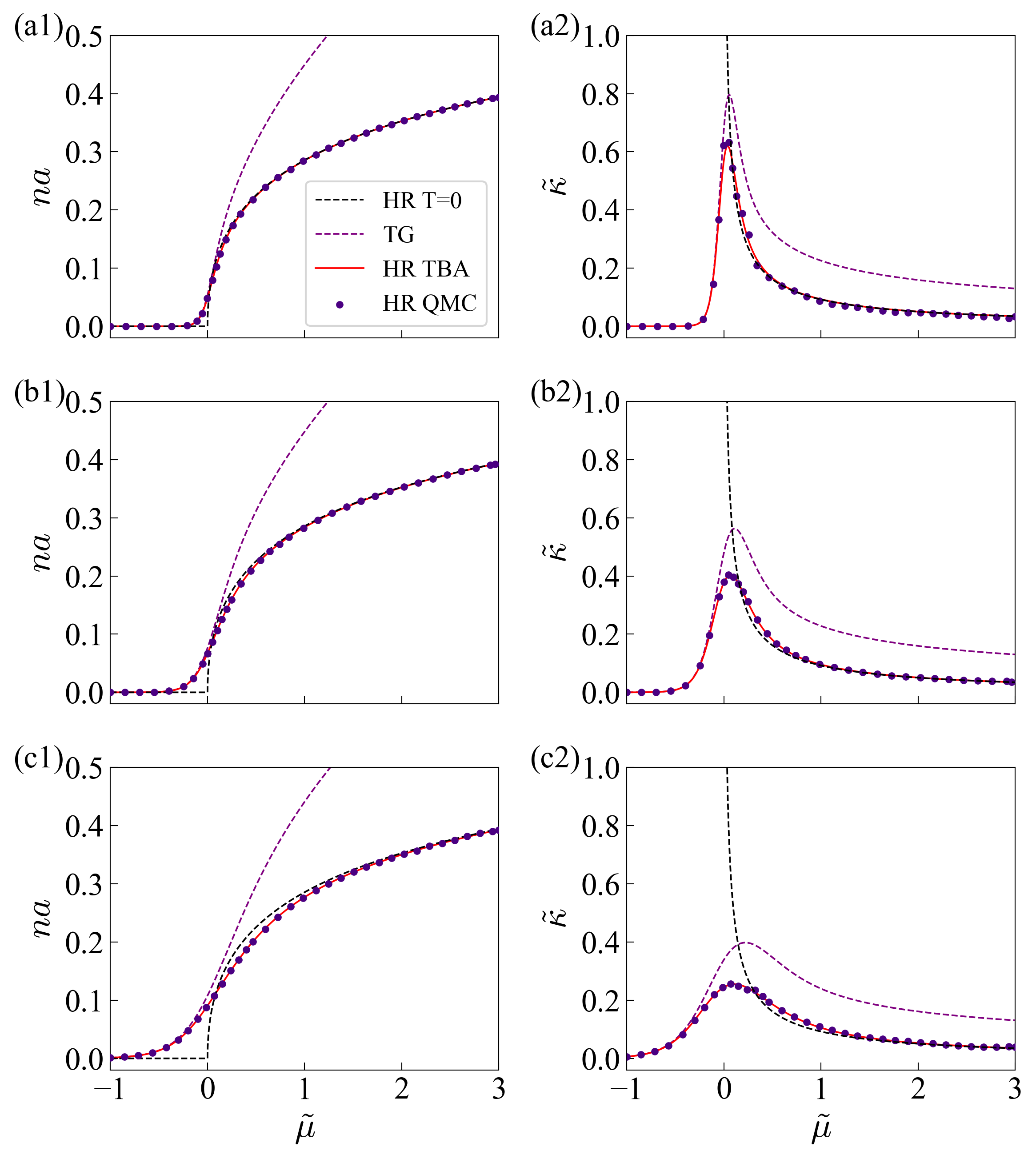}
\caption{
Thermodynamics of HRs.
(a)~Rescaled density, $na$, and (b)~compressibility, $\tilde{\kappa}=\hbar^2\kappa/ma$, versus chemical potential, $\tilde{\mu}=\mu ma^2/\hbar^2$, for the HR and TG models.
For HRs, we show TBA predictions (solid red line) and QMC results (purple disks) at $(ma^2\kB /\hbar^2) T=0.05, 0.1,$ and $0.2$ from top to bottom, as well as BA predictions at $T=0$ (dashed black line).
For TG gas, we show TBA predictions (dashed purple line). Errorbars are smaller than the size of markers and width of lines.}
\label{fig: eqnofstate}
\end{figure}
Figure~\ref{fig: eqnofstate} shows the density $na$ in the left column and the compressibility $\tilde{\kappa}=\kappa \hbar^2/ma$ in the right column versus chemical potential for various temperatures.
QMC results are shown for the HRs (purple disks) while TBA predictions are shown for HRs [solid red lines, Eqs.~(\ref{eq: density}) and (\ref{eq: compressibility})] and for the TG model (dashed purple line).
The thermodynamics of the TG gas is obtained by plugging $a=0$ into the Yang-Yang equation~(\ref{eq:yang-yang}), which directly yields $\epsilon(k)=-\mu + \hbar^2 k^2 / 2m$. Then we compute the density and the compressibility from Eqs.~(\ref{eq: density}) and (\ref{eq: compressibility}), respectively.
To circumvent the singularity at $a=0$, we compute $\chi_2/a$ and $\chi_3/a$.
Note that for the TG gas, the density $n$ is finite but $n\aHR$ vanishes since $\aHR=0$. Hence,
in Fig.~\ref{fig: eqnofstate}, the plotted data for the TG gas is $n\aHR$ with $\aHR$ the radius of the HRs considered in the same figure.
The zero-temperature behavior, predicted by BA, is also shown for the HR model [dashed black lines, from Eq.~(\ref{eq: mu})].
QMC results show excellent agreement with TBA across all chemical potentials and finite temperatures, confirming the TBA predictions.
Moreover, the scaling of QMC results for decreasing temperatures confirm the $T=0$ behavior predicted by BA.
Qualitatively similar results are found for all temperatures except for $T=0$.

At zero temperature, the density exhibits a sharp edge, $ n \propto \sqrt{\mu} $, consistent with Eq.~(\ref{eq: mu}) in the limit $n \rightarrow 0$.
This yields a diverging compressibility $ \kappa \propto 1 / \sqrt{\mu} $ around $ \mu \simeq 0 $.
The latter is associated to a vacuum-to-TLL-liquid quantum phase transition~\cite{sachdev2001}. Using Bose-Fermi mapping for the dilute gas, it may be understood as a van Hove singularity,
which is induced by a diverging density of states at zero energy~\cite{ashcroft1976solid}.
Finite-temperature effects smooth out this sharp edge of the density curves and eliminate the compressibility divergence.
The latter is replaced at finite temperature by a maximum at finite chemical potential, which gets sharper as the temperature decreases.
Comparing these results with those for the TG model,
we find that both models exhibit qualitatively similar behaviors but significant quantitative differences.
As expected, the TG and HR models agree in the low-density limit $n\aHR \ll 1$. However, significant deviations occur as the chemical potential increases due to the finite-separation impenetrability of the HR model. 
As a result, the compressibility of HRs is generically lower than that of the TG gas.
Moreover, results not shown here indicate that these deviations are stronger when the temperature increases owing to the different excitation spectra of the HR and TG models.
Note that, as anticipated, no simple proportional relation between the HR and TG models exist because the rescaling of energies and chemical potentials are different and depend on density and temperature.

\section{Correlation functions}
\label{sec: correlations}
Our results allow us to predict (up to numerical prefactors) the behavior of correlation functions assuming TLL behavior.
For instance, using mode expansions of the TLL fields,
the asymptotic one-body correlation function $g_1(x) = \langle \psi^\dagger(x) \psi(0) \rangle / n$ may be written, for $x \gg n^{-1}$, as~\cite{haldane1981}
\begin{equation}
g_1(x) = \sum_{m \geq 0} B_m\frac{\cos(2\pi n m x)}{|nd(x)|^{2m^2 K+1/2K}} , \label{eq:lut_g1}
\end{equation}
where $K$ is the TLL parameter
and
$d(x)=\frac{L_T}{\pi} \sinh\left(\frac{\pi x}{L_T} \right)$ is the thermal chord distance,
with $L_T = \hbar u/\kB T$ the thermal length and $u$ the speed of sound.
A similar formula is found for the pair correlation function $g_2(x) = \langle \psi^\dagger(0) \psi^\dagger(x) \psi(x) \psi(0) \rangle / n^2$, which reads as
\begin{equation} \label{eq:lut_g2}
g_2(x) = 1 - \frac{2K}{|2\pi nd(x)|^2} + \sum_{m \geq 1} A_m \frac{\cos(2\pi nmx)}{|nd(x)|^{2m^2K}}.
\end{equation}
For HRs, the speed of sound may be computed from the exact dispersion relation~(\ref{eq:e(p)_HR}), which yields
\begin{equation} \label{eq:luttinger_u}
u = \left. \frac{\D \varepsilon}{\hbar \D q} \right|_{q=0} = \frac{\pi \hbar n}{m(1-n\aHR)^2},
\end{equation}
and the TLL parameter is given by~\cite{cazalilla2004bosonizing}
\begin{equation} \label{eq: luttinger K} 
K = {\hbar\kF}/{mu} = (1-n\aHR)^2.
\end{equation}
The nonuniversal (model-dependent) coefficients $B_m$ and $A_m$ are, however, not determined by the TLL theory.

\begin{figure}[tbp]
\centering
\includegraphics[width=1\columnwidth]{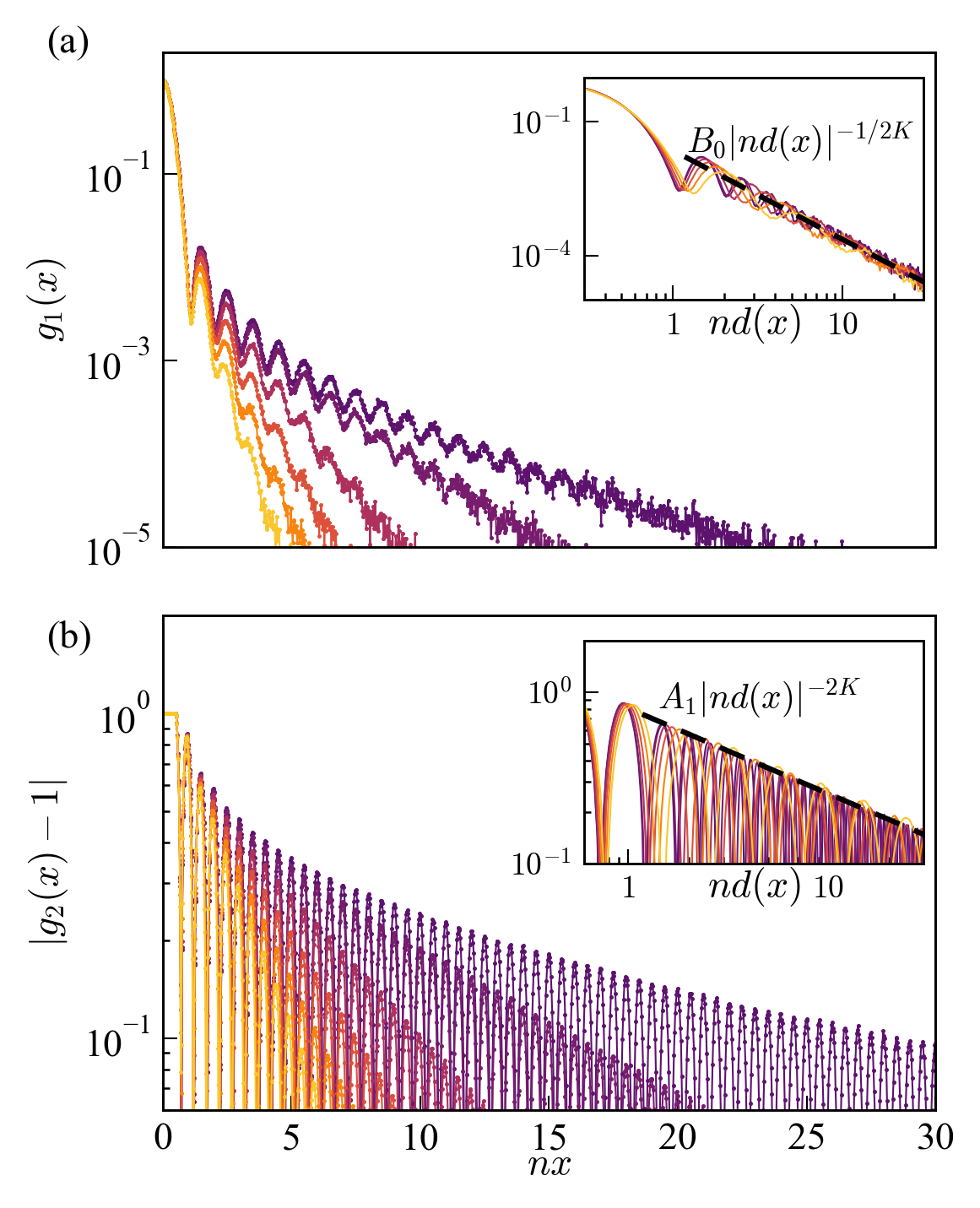}
\caption{
Correlations in the HR model.
(a)~One-body correlation function $g_1(x)$ and (b)~Pair correlation function $g_2(x)$ for $n\aHR=0.5$ and various temperatures, in semi-log scale.
Solid lines are QMC results with temperatures $(\kB m\aHR^2/\hbar^2) T=0.1, 0.2, 0.4, 0.6, 0.8, 1.0$ from top (dark purple) to bottom (light yellow),
corresponding to $nL_T = 31.4, 15.7, 7.86, 5.24, 3.93, 3.15$, respectively.
The insets show the same data as in the main panels but plotted versus the thermal chord distance $d(x)$ in a log-log scale, and the dashed lines indicate the leading term of TLL predictions, Eqs.~(\ref{eq:lut_g1}) and (\ref{eq:lut_g2}), where $B_0$ and $A_1$ are fitted parameters and $K$ is computed using Eq.~(\ref{eq: luttinger K}).
}
\label{fig: correlation functions}
\end{figure}
Figure~\ref{fig: correlation functions} shows the complete correlation functions $g_1(x)$ and $|g_2(x)-1|$ as computed using QMC for \( n\aHR=0.5 \) and increasing temperatures (from top to bottom).
We plot $|g_2(x)-1|$ instead of $g_2(x)$, to highlight the oscillations.
The main panels show the correlation functions plotted in semi-log scale versus the distance $x$ and the insets show the same data plotted in log-log scale versus the thermal chord distance $d(x)$.
The plateau visible in the main panel of~(b) for $nx<na=0.5$, where $g_2(x)=0$ ($|g_2(x)-1|=1$), is a result of the fully impenetrable nature of HRs.
The asymptotic behaviors are consistent with Eqs.~(\ref{eq:lut_g1}) and (\ref{eq:lut_g2}).
In particular, both $g_1(x)$ and $g_2(x)$ exhibit oscillations versus $x$ with period $1/n$, irrespective of temperature, and
decay stronger with increasing temperature, consistently with $L_T \propto 1/T$ (see main panels in Fig.~\ref{fig: correlation functions}).
For $x\ll L_T$, we find power-law decay reminiscent of the universal zero-temperature TLL behavior, while
for $x \gg L_T$, we have $d(x)\sim \exp(\pi x/L_T)$ and exponential decay is found~(the values of $L_T$ are given in the caption of Fig.~\ref{fig: correlation functions}).
Such a crossover is found in both $g_1(x)$ and $|g_2(x)-1|$. 
Note that due to a relative decay factor $|nd(x)|^{-2K}$, the dominating $m=1$ oscillations of $g_1(x)$ are suppressed with respect to the $m=0$ term in Eq.~(\ref{eq:lut_g1}) at large distance and high temperature.
More precisely, universal algebraic decay (up to oscillations) with respect to $d(x)$ is expected.
This is verified in the inset of Fig.~\ref{fig: correlation functions}(a), which shows $g_1(x)$ versus $d(x)$ in log-log scale for the same temperatures as in the main panel,
and the dashed line denotes the leading ($m=0$) term of Eq.~(\ref{eq:lut_g1}), with $K$ as given by Eq.~(\ref{eq: luttinger K}) and $B_0$ as a fitting parameter.
We find all curves collapse up to the oscillations, in agreement with the TLL predictions.
Similarly, the inset of Fig.~\ref{fig: correlation functions}(b) shows $|g_2(x)-1|$ versus $d(x)$ in log-log scale.
The dominating term is the absolute value of the $m=1$ oscillation.
Since $K<1$ for HRs, a universal power-law upper bound $A_1\vert n d(x) \vert^{-2K}$ is observed in this case, as indicated by the dashed line, which once again confirms the TLL predictions.

\begin{figure}[tbp]
\centering
\includegraphics[width=1\columnwidth]{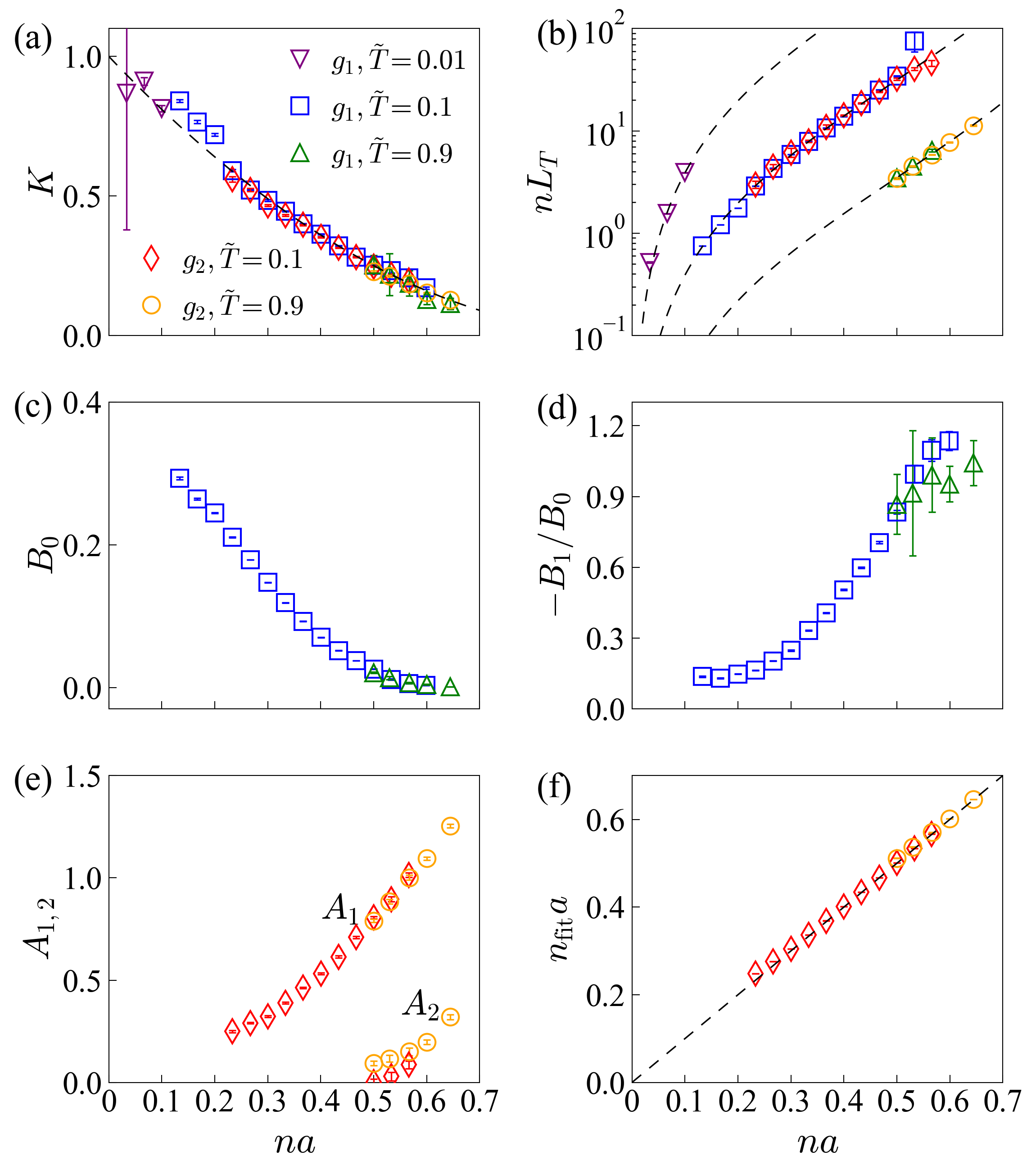}
\caption{
Results of correlation function fits. We fit the TLL predictions, Eqs.~(\ref{eq:lut_g1}) and (\ref{eq:lut_g2}) to QMC results,
for $\tilde{T}=m\aHR^2\kB T/\hbar^2=0.01$ (purple triangles) $0.1$ (blue squares and red diamonds) and $\tilde{T}=0.9$ (green triangles and orange circles).
For $g_1(x)$, we take $K$, $L_T$, $B_0$, and $B_1$ as fitting parameters, and for $g_2(x)$, $K$, $L_T$, $n$, and $A_1$.
Errorbars are found to be smaller than the markers.
The black dashed lines indicate TLL theory predictions.
In~(b) the three dashed lines indicate the thermal lengths for the considered temperatures.}
\label{fig:correlations_fits}
\end{figure}
To further asses the TLL behavior, we now proceed with fits of Eqs.~(\ref{eq:lut_g1}) and (\ref{eq:lut_g2}), up to the $m=1$ terms, to the QMC data for various values of $n\aHR$.
This allows us to extract $K$, $L_T$, and the coefficients $B_0$, $B_1$, and $A_1$. 
Furthermore, for $g_2(x)$, we fit the density $n$ from the frequency of oscillations.
We use a Levenberg-Marquardt least-square approach~\cite{more2006levenberg} in two steps.
In the first step, we fit $g_1(x)$ and obtain estimates for $K$, $B_0$, and $B_1$.
The thermal length $L_T$ is, however, poorly estimated because thermal effects are only significant at large enough distance, $x \gtrsim L_T$, where the value of the correlation function is small and does not contribute much to the fit.
To enhance the weight of the tails, we perform a second fit, now of $\ln [g_1(x)]$, using the previous estimates for $K$, $B_0$, and $B_1$, and fitting the parameter $L_T$ only.
The speed of sound $u$ can then be extracted knowing the temperature, here $\tilde{T}=m\aHR^2\kB T/\hbar^2=0.01$ for $na\leq 0.1$, $\tilde{T}=0.1$ for $0.13\leq n\aHR \leq 0.6$ and $\tilde{T}=0.9$ for $0.5\leq na\leq 0.7$, respectively.
These temperatures are chosen so that the system always remains in the TLL regime, \ie~in a regime in which the nonlinear part of the excitation spectrum can be neglected.
This condition is fulfilled when $\kB T$ does not exceed $\hbar^2 \pi^2 n^2 / 2m(1-na)^2$.
Due to convergence issues, the extension to $na\gtrsim 0.6$ is more feasible for the highest temperature.
A similar analysis is carried out for $g_2(x)$.
In the first step, we fit $g_2(x)$ to obtain estimates of $K$, $n$, and $A_1$.
For $na\geq 0.5$, we also include $A_2$ into the fits since this term becomes significant.
In the second step, we fit all local maxima of $\ln|g_2(x)-1|$ to extract $L_T$ in order to exclude the points with $g_2(x) \approx 1$ , which would otherwise yield large values of $\ln |g_2(x)-1|$ and strongly affect the fits.
For all cases, we find excellent agreement between the fitted curves and the QMC data,
see examples
in Appendix~\ref{appendix:fits}.

The results of the fits are shown in Fig.~\ref{fig:correlations_fits},
with errorbars estimated by the residual sum of squares.
Purple, blue, and green data correspond to fits of $g_1$ for the three different temperatures, while red and orange data correspond to fits of $g_2$ for the higher two temperatures.
We find very good agreement with TLL theory for the parameter $K$ obtained by fitting $g_1(x)$, except for a few points with $n\aHR \lesssim 0.2$, where the agreement is worse, but still fair with deviations less than $13\%$.
The deviation at small values of $n\aHR$ is due to the fact that dilute density leads to weak oscillations, which are further suppressed exponentially by finite temperature.
The correlation function $g_1(x)$ approaches a pure exponential decay proportional to $\exp(-\pi x/2KL_T)$. 
The latter depends on $KL_T$, rather than $K$ and $L_T$ independently, leading to poor fitting of $K$.
For $n\aHR\lesssim 0.1$, the oscillations are actually invisible, and the exponential decay renders the contribution of $B_1$ negligible in the fitting. 
It results in large uncertainties and the corresponding fits of $B_0$ and $B_1/B_0$ are not shown.
Nevertheless, the behaviours of our results are consistent with the TG limit for $n\aHR \to 0$,
with $K\to 1$, $nL_T \sim (n\aHR)^2$ [note the semi-log scale in Fig.~\ref{fig:correlations_fits}(b)], and the exact coefficients $B_0\approx 0.52$ and $-B_1/B_0\approx 1/8=0.125$~\cite{cazalilla2011}.
Similarly,
for $g_2$, the weak amplitude and rapid decay of the oscillations allow less than two oscillation periods to be resolved before they are overwhelmed by statistical fluctuations, making a reliable fit impossible for $na\leq 0.2$.
In the range $0.2\lesssim n\aHR$, the estimates for $K$ from $g_2(x)$ show good agreement with TLL theory, confirming the TLL theory predictions.
Moreover, in the overlap region of $\tilde{T}=0.1$ and $0.9$, $0.5 \lesssim na \lesssim 0.6$, all fits agree well, confirming that the parameter $K$ and all fitted coefficients are independent of $T$, thus validating the TLL predictions again.

The fits for $L_T$, as obtained in the second fit step, are also in good agreement with the TLL prediction, except for $n\aHR \gtrsim 0.5 \textrm{-} 0.6$.
In this regime, the small value of $K$ leads to fast decay of $g_1$ [see Eq.~(\ref{eq:lut_g1})], hiding the thermal effects under statistical fluctuations.
Then, we cannot fit $g_1(x)$ for $L_T$ although the fits for $K$ obtained in the first fit still work well in this regime since they are not affected by the large value of $L_T$.
In contrast, the oscillations of $g_2(x)$ decay slower because the $2K$ exponent is smaller than the exponent $1/2K+2K$ of $g_1$ for $m=1$, especially for large values of $n\aHR$.
The valid range is larger against the QMC statistical fluctuations.
Thus, we are able to fit $L_T$ from $g_2$ in a larger range to obtain good estimates of $L_T$.

The quality of the fits for $K$ and $L_T$, which yield results in very good agreement with the theoretical predictions, validates the reliability of the extracted coefficients $B_0$, $B_1$, $A_1$, and $A_2$.
The increase in $B_1/B_0$, $A_1$, and $A_2$ indicates a enhancement of HRs solidification-like behavior.
Furthermore, fits for $n$ are shown in Fig.~\ref{fig:correlations_fits}(f).
We find that $n_\mathrm{fit}$ agrees with $n$ perfectly, validating the predicted period of oscillation.

\begin{figure}[tbp]
\centering
\includegraphics[width=\columnwidth]{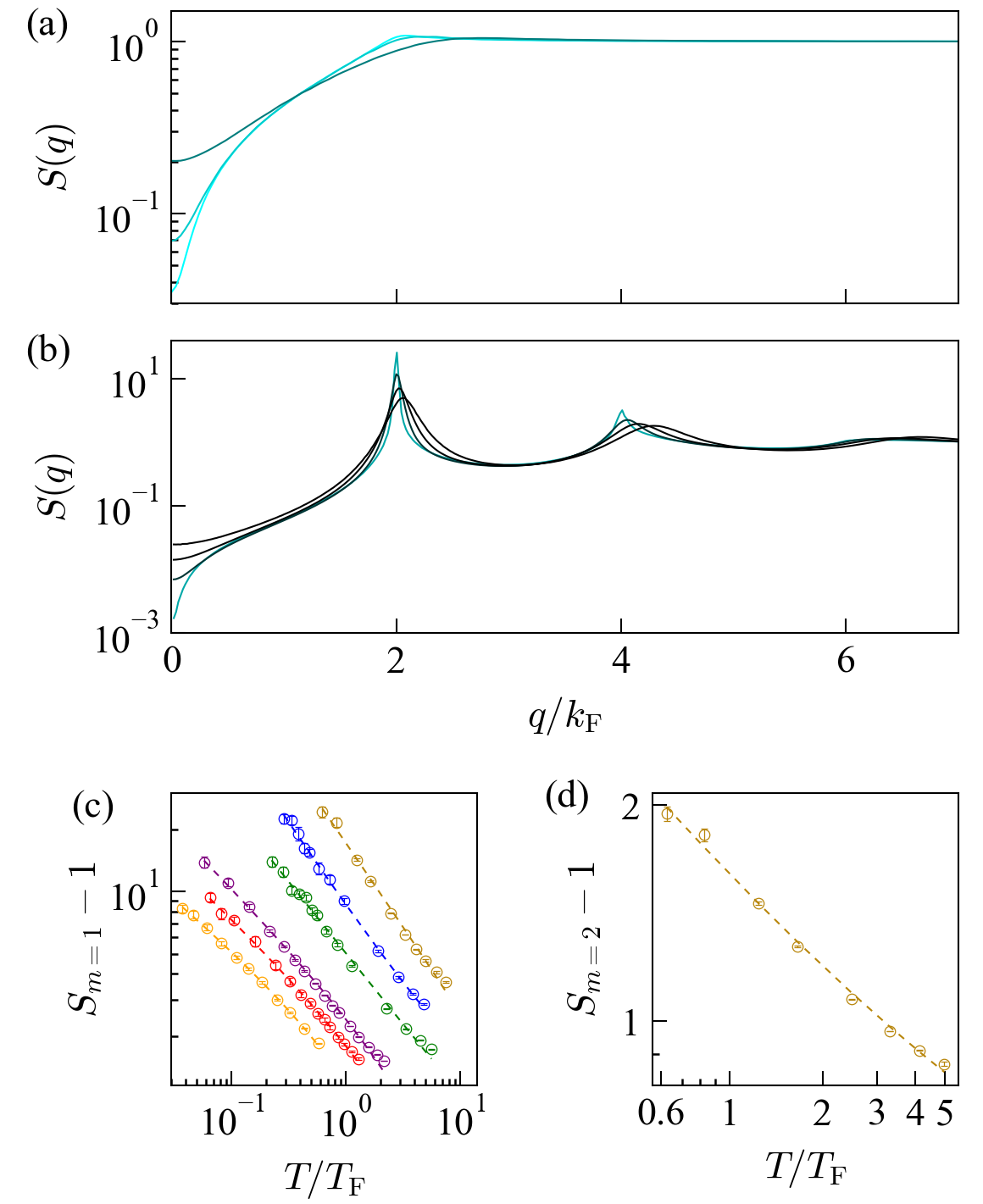}
\caption{
Finite-temperature static structure factor
$S(q)$ obtained from QMC computations at (a)~$na=0.1$ and temperatures $T/\TF =0.101, 0.203$ and $0.608$, and (b)~at $n\aHR=0.7$ and $T/\TF=0.339$, $1.24$, $1.65$, $3.31$, $6.20$, plotted from light to dark colors, where $\TF = \EF / \kB$ is the Fermi temperature.
The lower panels show the amplitude of the first peaks, (c)~$S_{m=1}$ for various values of $na$ and (d)~$S_{m=2}$ for $na=0.7$, versus temperature in log-log scale.
Dashed lines are fitted power-law scalings, shifted by a constant $C$ as given in Eq.~(\ref{eq:SmScaling}), with fitted parameters and corresponding $na$ shown in Table~\ref{tab:fits_S(k)}.
}
\label{fig:Sk}
\end{figure}
Finally, we turn to the static structure factor $S(q)=\frac{1}{N}[\langle \rho_q \rho_{-q} \rangle - |\langle \rho_q \rangle|^2 ]$ where $\rho_q=\sum_{j=1}^N e^{-iqx_j}$.
It is related to the pair correlation function through a Fourier transform,
\begin{equation} \label{eq:Sk_g2_relation}
S(q) = 1 + n\int_{-\infty}^{\infty} \D x [g_2(x)-1] e^{iqx}.
\end{equation}
For HRs at zero temperature, it was shown earlier that non-analytical diverging peaks of $S(q)$ appear at $q=2m\kF$ when $2m^2K<1$, arising from the $m$-th summation term in Eq.~(\ref{eq:lut_g2})~\cite{mazzanti2008ground,astrakharchik2014luttinger,motta2016dynamical}.
At finite temperature, since the correlations turn to exponential decay for $x\gg L_T$, the behavior of the $m$-th peak, $S_m = S(q \simeq 2m\kF)$,
changes accordingly. Inserting Eq.~(\ref{eq:lut_g2}) into Eq.~(\ref{eq:Sk_g2_relation}), we find power-law scaling with temperature,
\begin{equation} \label{eq:SmScaling}
S_m - 1 \approx \frac{A_m \Gamma\left(\frac{1}{2} \! - \! m^2K\right)\Gamma\left( m^2K \right)}{2\sqrt{\pi}} \! \left(\frac{n L_T}{\pi}\right)^{1-2m^2K} \! + C,
\end{equation}
where
we collect the contributions of the terms other than $q=2m\kF$ into a constant $C$.
This leads to the scaling with thermal length $S_m - 1 \propto L_T^{1-2m^2K}$.
Note that, for a finite-size system with size $L$, at zero-temperature, the thermal length is replaced by the system size $L$, which yields the scaling $S_m(T=0) - 1 \propto L^{1-2m^2K}$, consistent with the scaling with $N$ found in Ref.~\cite{mazzanti2008ground}.
It shows that $S_m(T=0)$ scales up with increasing $L$ for $1-2m^2K>0$.
At finite temperature, the non-analytical peak present at zero temperature is replaced by a local maximum as indicated by Eq.~(\ref{eq:SmScaling}), the height of which scales down with temperature as $S_m-1 \propto 1/T^{1-2m^2K}$ since $L_T \propto1/T$.
The $m$-th peak emerges when $na > 1-1/\sqrt{2m^2}$, which corresponds to
$na>0.293$ for the first peak ($m=1$) and to $na>0.646$ for the second peak ($m=2$).
Such peaks indicate the existence of a quasisolid regime at high densities.
The observed diverging $S(q)$ signals the onset of quasi-solid behavior, implying the enhancement of zero-energy modes at finite-momentum.
This is reminiscent of the supersolids in higher dimensions~\cite{saccani2012excitation,tanzi2019supersolid}, where the softened roton gives rise to true long-range order.
However, due to the reduced dimensionality here, although the ordering is significantly enhanced, it does not develop into true long-range order, and the system continues to fall within the TLL regime.
In contrast, such peaks do not appear in the Lieb-Liniger model.

The  static structure factors at $n\aHR=0.1$ and $n\aHR=0.7$ for various temperatures are shown in Fig.~\ref{fig:Sk}(a) and (b), respectively, where lighter colors indicate lower temperatures.
For $n\aHR=0.1$, no peaks nor divergences are observed, consistently with the fact that $1-2K<0$.
In contrast, for $na=0.7$, the first and second sharp peaks around $q=2\kF$ and $q=4\kF$, respectively, are clearly visible, consistently with the fact that $1-2m^2K>0$ for $m=1$ and $m=2$.
The peaks exhibit Lorentzian-like forms, with amplitudes that decrease with increasing temperature, implying that the quasi-solid ordering is weakened by thermal fluctuations.
Meanwhile, the position of the peak slightly shifts towards larger momenta.
This is consistent the shift of the oscillating period of $g_2(x)$ observed at these temperatures (not shown here).
The power law scaling of the peak amplitudes $S_m$, Eq.~(\ref{eq:SmScaling}), is confirmed in Figs.~\ref{fig:Sk}(c) and (d) for $m=1$ and $m=2$, respectively.
Both clearly show linear scalings in log-log scale, consistent with Eq.~(\ref{eq:SmScaling}). 
More precisely, we fit Eq.~(\ref{eq:SmScaling}) to the data, with $K,\, A_m$, and $C$ as fitting parameters. 
The QMC data for $S_m$ (dots), together with the fits (dashed lines) are shown for various values of $n\aHR$, with fitted parameters given in Table~\ref{tab:fits_S(k)},
\begin{table}[]
\caption{Fits for $S_m(T)$.}
\label{tab:fits_S(k)}
\begin{tabularx}{\columnwidth}{X@{\hskip -0.5cm}X@{\hskip -0.5cm}XX@{\hskip 0.5cm}X@{\hskip 0.5cm}X}
\toprule
$na$ & m & $K_\mathrm{BA}$ & $K_\mathrm{fit}$  & $A_m$ & $C$ \\
\midrule
0.47 & 1 & 0.2844 & 0.28(2) & 0.70(12) & -1.1(5) \\
0.5 & 1 & 0.25 & 0.25(2) & 0.76(14) & -0.7(4)  \\
0.53 & 1 & 0.2209 & 0.227(12) & 0.89(10) & -0.70(3) \\
0.6 & 1 & 0.16 & 0.15(3) & 0.9(3) & 0.1(4) \\
0.65 & 1 & 0.1225 & 0.09(4) & 0.6(6) & 0.5(9) \\
0.7 & 1 & 0.09 & 0.10(5) & 1.6(17) & -0.2(15) \\
0.7 & 2 & 0.09 & 0.06(2) & 0.16(10) & 0.3(2) \\
\bottomrule
\end{tabularx}
\end{table}
where $K_\mathrm{BA}=(1-na)^2$ is the BA prediction.
We performed the fits in the range for $na\geq 0.47$, where the first peak is sufficiently large.
We find good agreement for the $m=1$ peaks, with relative discrepancy of $K_\mathrm{fit}$ less than 6\% for $na\leq 0.6$.
Furthermore, the value of $A_1$ at $na=0.5$ agrees with the result of $g_2(x)$, and $A_1\approx 1$ at $na=0.6$ agrees with the result found at zero temperature in Ref.~\cite{mazzanti2008ground}.
For $na\gtrsim 0.65$, the large density prevents QMC from converging at relatively lower temperatures and thus it is difficult to reach larger values of $S_m$~\cite{ceperley1995}.
Finally, the $m=2$ peak is identified at $na=0.7$ and the scaling relation is observed, see Fig.~\ref{fig:Sk}(d).
Although the fit is worse than those of $S_{m=1}$, it still has a correct order of magnitude.
In all simulations, the particle number is larger than $200$.
These results further confirm the TLL behavior of HRs at finite temperature.

\section{Conclusion and discussion}\label{sec:conclusions}
In conclusion, we reported the investigation of the thermodynamics of the 1D HR gas.
This model shows significant deviations compared to the Lieb-Liniger model, which corresponds to contact interactions, in regimes with relatively high density.
By comparing the TBA solution and universal TLL predictions with QMC calculations for thermodynamic properties and correlation functions,
we found excellent agreement.
Our findings highlight the universal TLL behavior of the HR model across a wide range of parameters, as evidenced by correlation functions and static structure factors.
We have shown that the parameters of the TLL model $K$ and $u$ can be extracted from fits of correlation functions.
More precisely, we obtained the thermal length $L_T = \hbar u / \kB T$, from which the speed of sound $u$ can be deduced given the temperature $T$. Conversely, the same approach can be used to infer the temperature from the analytical value of the speed of sound. This makes the approach useful for thermometry in experiments.
Furthermore, the numerical coefficients $B_0$, $B_1$, and $A_1$, which are unknown in the TLL approach, can also be extracted.
An interesting direction for future studies would be to compute the dynamical structure factor at finite temperatures, which is of experimental relevance for instance in ultracold-atom systems.

Rydberg atom systems~\cite{browaeys2016,browaeys2020} may offer a promising platform to implement approximations of the HR model we have discussed in this paper.
The bare van der Waals interactions already provides a sharp decay of interactions as the sixth power of the interatomic distance $r$, showing hard core behavior of correlation functions on lattices~\cite{labuhn2016tunable}. This can be further improved using Rydberg dressing of the ground state, with effective interactions showing a plateau with a sharp $1/r^6$ decay~\cite{pupillo2010,honer2010,johnson2010,plodzien2017} or even sharper via the electromagnetically induced transparency~\cite{gaul2016resonant,helmrich2016two},
and controlling of the width and height of the plateau controlled via the intensity and detuning of the coupling lasers.
The excitation spectrum could then be measured using for instance quench spectroscopy~\cite{menu2018,menu2023,villa2019,villa2020,villa2021a,villa2021b}
as recently implemented in Rydberg atom experiments~\cite{chen2025}.
Furthermore, thermodynamic properties may be studied by varying the interaction range $\aHR$ and/or the temperature by controlling the cooling ramp.
Further investigation is deserved to assess the accuracy of such implementations in realizing the HR model.

We thank Antoine~Browaeys, Jacopo De~Nardis, Dean Johnstone, and Zoran Ristivojevic for fruitful discussions. We acknowledge the CPHT computer team for valuable support.
This research was supported by
the Agence Nationale de la Recherche (projects ANR-CMAQ-002 and ANR-23-PETQ-0002 France 2030),
the IPParis Doctoral School, 
and HPC/AI resources from GENCI-TGCC (Grant 2023-A0110510300) using the ALPS scheduler library and statistical analysis tools~\cite{troyer1998,ALPS2011}.

\appendix


\section{Numerical solution of the Yang-Yang equation}
\label{appendix:dressed-energy}
To solve the Yang-Yang equation, Eq.~(\ref{eq:yang-yang}), for HRs, we first rewrite it as
\begin{equation}
\epsilon(k)  = -\mu + \frac{\hbar^2 k^2}{2m} + \kB T \aHR \int \frac{\D k'}{2\pi} \ln \left[ 1 + e^{-\epsilon(k') / \kB T} \right],
\label{eq:yang-yang-appendix}
\end{equation}
using $K(k-k')=\aHR$.
Note that the integral depends on the yet unknown form of the function $\epsilon(k)$ but, contrary to the Lieb-Liniger case, is a constant, independent of $k$.
It is then fruitful to rewrite it as
\begin{equation}
\epsilon(k)  = -\mu + \frac{\hbar^2 k^2}{2m} + \kB T \chi_1(\mu,T),
\label{eq:yang-yang-appendixBIS}
\end{equation}
with
\begin{equation}
\chi_1(\mu,T) = \aHR \int \frac{\D k'}{2\pi} \ln \left[ 1 + e^{-\epsilon(k') / \kB T} \right],
\label{eq:yang-yang-appendix.chi1}
\end{equation}
which explicitly shows that $\epsilon(k)$ is a quadratic function of $k$.
Inserting Eq.~(\ref{eq:yang-yang-appendixBIS}) into Eq.~(\ref{eq:yang-yang-appendix.chi1}), we then find the self-consistent equation
\begin{equation}
\chi_1(\mu,T) = -\frac{\aHR}{\lambda_T} {\rm Li}_{3/2}\left(-\e^{\mu/\kB T - \chi_1}\right),
\label{eq:yang-yang-appendix.simple}
\end{equation}
where ${\rm Li}_{3/2}$ is the polylogarithm function and $\lambda_T=\sqrt{2\pi\hbar^2/m\kB T}$ is the thermal de Broglie wavelength.
The right-hand-side term is a strictly decreasing function of $\chi_1$, from a positive value for $\chi_1=0$ to zero for $\chi_1 \rightarrow \infty$.
Equation~(\ref{eq:yang-yang-appendix.simple}) has thus a unique solution for each value of $\mu$ and $T$, which is easily found using an iterative Newton method.
Inserting the result into Eq.~(\ref{eq:yang-yang-appendixBIS}), we then find $\epsilon(k)$.

\begin{figure}[t!]
\centering
\includegraphics[width=\linewidth]{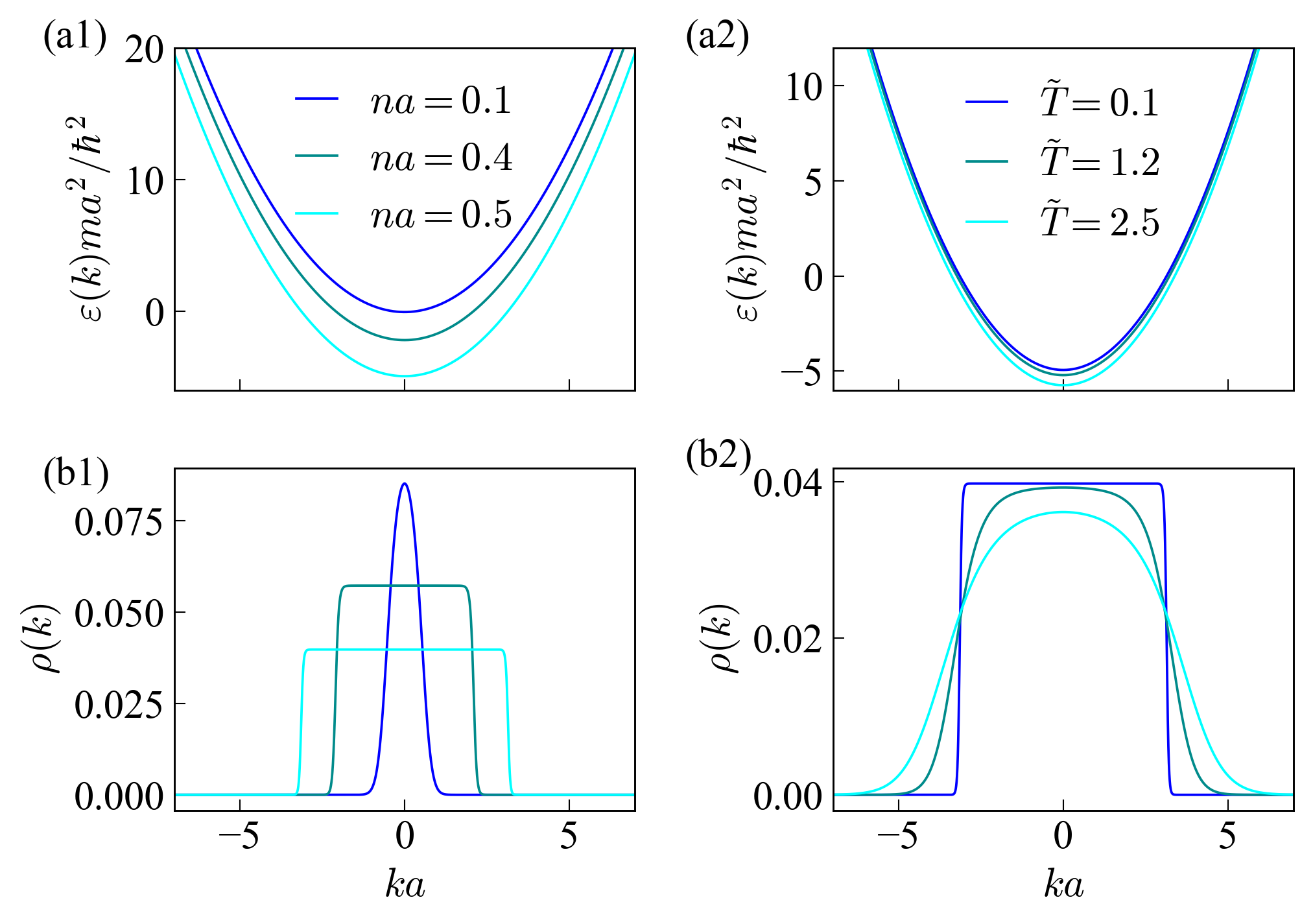}
\caption{(a) Dressed energy and (b) density of quasi-momenta. Left: Various values of $na$ for $T=0.1 \hbar^2 / ma^2\kB$. Right: Various temperatures for $na=0.5$}
\label{fig:dressed-energy_appendix}
\end{figure}

After solving the Yang-Yang equation for $\epsilon(k)$, we can compute a number of quantities, including
the quasi-momentum density, given by $\rho(k) = \frac{1-na}{2\pi} \frac{1}{1 + \e^{\epsilon(k)/\kB T } }$,
see Eq.~(\ref{eq:App.rhok}) in Appendix~\ref{appendix:density-compressibility}.
Examples of the dressed energy $\epsilon(k)$ and the corresponding $\rho(k)$ are shown in Fig.~\ref{fig:dressed-energy_appendix}. 
As expected, we find quadratic behavior of $\epsilon(k)$ as a function of $k$, and the Fermi-distribution-like density of quasi-momenta. 
Note that $\rho(k)$ is not the momentum distribution, which equals the Fourier transform of the off-diagonal one-body density matrix.

\section{Derivation of the density and compressibility}
\label{appendix:density-compressibility}
To find the density, we first write
\begin{equation}
\rho_h(k) = \e^{\epsilon(k)/ \kB T}\rho(k),
\end{equation}
using Eq.~(\ref{eq:constraint_rho_rhoh_epsilon}).
Since $\calK(k)=\aHR$ and $n=\int dk \rho(k)$, Eq.~(\ref{eq:constraint_rho_rhoh}) reduces
$
\rho(k) + \rho_h(k) = \frac{1-n\aHR}{2\pi},
$
and one finds
\begin{equation}\label{eq:App.rhok}
\rho(k) = \frac{1-n\aHR}{2\pi}  \frac{1}{1 + \e^{\epsilon(k)/ \kB T}}.
\end{equation}
Then, integrating over $k$ on both sides, one finds
\begin{equation}\label{eq:App.density}
n = (1-n\aHR) \frac{\chi_2(\mu, T)}{a},
\end{equation}
with $\chi_2(\mu,T) = \aHR \inftyint \frac{\D k}{2\pi} \frac{1}{1 + \e^{\epsilon(k)/\kB T}}$.
Solving for $n$, one then finds Eq.~(\ref{eq: density}). 

The compressibility is found by differentiating Eq.~(\ref{eq: density}) with respect to $\mu$, which yields
\begin{equation} \label{eq:dn_dmu}
\kappa = \left( \frac{\pa n}{\pa \mu} \right)_T = \frac{1}{\aHR} \frac{1}{(1 + \chi_2)^2}  \frac{\pa \chi_2(\mu, T)}{\pa \mu}.
\end{equation}
Besides, the definition of $\chi_2(\mu, T)$ leads to
\begin{equation} \label{eq:dchi2_dmu}
\frac{\pa \chi_2(\mu, T)}{\pa \mu} = - \frac{\aHR}{\kB T} \inftyint \frac{\D k}{2\pi} \frac{\e^{\epsilon(k) / \kB T}}{\left[ 1 + \e^{\epsilon(k) / \kB T} \right]^2} \frac{\pa \epsilon(k)}{\pa \mu}.
\end{equation}
The Yang-Yang equation for HRs, Eq.~(\ref{eq:yang-yang-appendixBIS}), yields
\begin{equation} \label{eq:depsilon_dmu}
\frac{\pa \epsilon(k)}{\pa \mu} = -1 + \kB T \frac{\pa \chi_1}{\pa \mu},
\end{equation}
which is thus independent of $k$.
Using Eq.~(\ref{eq:yang-yang-appendix.chi1}), we then find
\begin{eqnarray} 
\frac{\pa \epsilon(k)}{\pa \mu}
& = & -1 - \aHR \inftyint \frac{\D k'}{2\pi} \frac{\e^{-\epsilon(k') / \kB T}}{1 + \e^{-\epsilon(k') / \kB T}} \frac{\pa \epsilon(k)}{\pa \mu}
\label{eq:depsilon_dmuBIS} \\
& = & -1 - \chi_2(\mu,T) \frac{\pa \epsilon(k)}{\pa \mu}
\nonumber
\end{eqnarray}
and, solving for ${\pa \epsilon(k)}/{\pa \mu}$, we find
\begin{eqnarray}\label{eq:appendix.depsilon}
\frac{\pa \epsilon(k)}{\pa \mu} = \frac{-1}{1 + \chi_2(\mu, T)}.
\end{eqnarray} 
Inserting Eq.~(\ref{eq:appendix.depsilon}) into Eq.~(\ref{eq:dchi2_dmu}), we get
\begin{equation} \label{eq:dchi2_dmuBIS}
\frac{\pa \chi_2(\mu, T)}{\pa \mu} = \frac{1}{\kB T} \times \frac{\chi_3(\mu,T)}{1 + \chi_2(\mu,T)},
\end{equation}
with
$\chi_3(\mu,T) = \aHR \inftyint \frac{\D k}{2\pi} \frac{ e^{\epsilon(k)/\kB T} }{ \left(1 + e^{\epsilon(k)/\kB T}\right)^2 }$.
Inserting this formula unto Eq.~(\ref{eq:dn_dmu}) and using Eq.~(\ref{eq:App.density}), we finally find Eq.~(\ref{eq: compressibility}).

\section{Propagator for quantum Monte Carlo}
\label{appendix:propagator}
\renewcommand{\theequation}{A.\arabic{equation}}
Here we derive the propagator for the QMC calculations.
The path-integral Monte Carlo formulation in continuous space involves slicing the imaginary time into many small slices with imaginary time step $\tau$.
In each slice, the pair interaction propagator, \ie~the probability amplitude for the relative separation to move from $x$ to $x'$, is given by~\cite{yan2015incorporating}
\begin{equation} \label{eq:app_propagator}
\begin{aligned}
\rho^{\mathrm{rel}}(x, x', \tau) = & \sum_n \psi_n^*(x) e^{-\tau E_n/\hbar} \psi_n(x') + \\ & \int_0^\infty \psi_k^*(x) e^{-\tau\hbar k^2/2m_{\mathrm{rel}}} \psi_k(x') \D k,
\end{aligned}
\end{equation}
where $m_{\mathrm{rel}} = m/2$ is the reduced mass,
$\psi_n(x)$ spans the set of bound states with energy $E_n$,
and $\psi_k(x)$ spans the set of scattering states,
for the reduced-particle Hamiltonian
\begin{equation}
H^{\mathrm{rel}} = -\frac{\hbar^2}{2m_{\mathrm{rel}}} \frac{\pa^2}{\pa x^2} + V(x).
\end{equation}
For HR interaction, there is no bound state and the scattering states are
\begin{equation} \label{eq:app_propagator_states}
\psi_k^{\mathrm{s,a}}(x) = \begin{cases}
\frac{1}{\sqrt{\pi}} \sin \left[ k(x-\aHR) \right], &\text{ for } x>\aHR, \\
\mp \frac{1}{\sqrt{\pi}} \sin \left[ k(x+\aHR) \right], &\text{ for } x < -\aHR, \\
0, &\text{ otherwise.}
\end{cases}
\end{equation}
The symmetric and anti-symmetric states are denoted by the superscripts ``s'' and ``a'', respectively. Substituting the expression of the scattering states, Eq.~(\ref{eq:app_propagator_states}), into Eq.~(\ref{eq:app_propagator}) we obtain
\begin{equation}
\rho(x,x',\tau) =  \left\{
\begin{aligned}
&\sqrt{\frac{m}{4\pi \tau \hbar^2}} \left( e^{-\frac{m(x-x')^2}{4\tau \hbar^2}} - e^{-\frac{m(|x+x'|-2\aHR)^2}{4\tau \hbar^2}} \right) \\
&\qquad\qquad \text{for } xx'> 0 \text{ and } |x|, |x'| > \aHR, \\
&0 \qquad\quad~~ \text{otherwise.}
\end{aligned} \right.
\end{equation}
This analytic expression is directly implemented in the QMC numerical code.

\section{Fits of correlation functions}
\label{appendix:fits}
\begin{figure}[h!]
\centering
\includegraphics[width=\columnwidth]{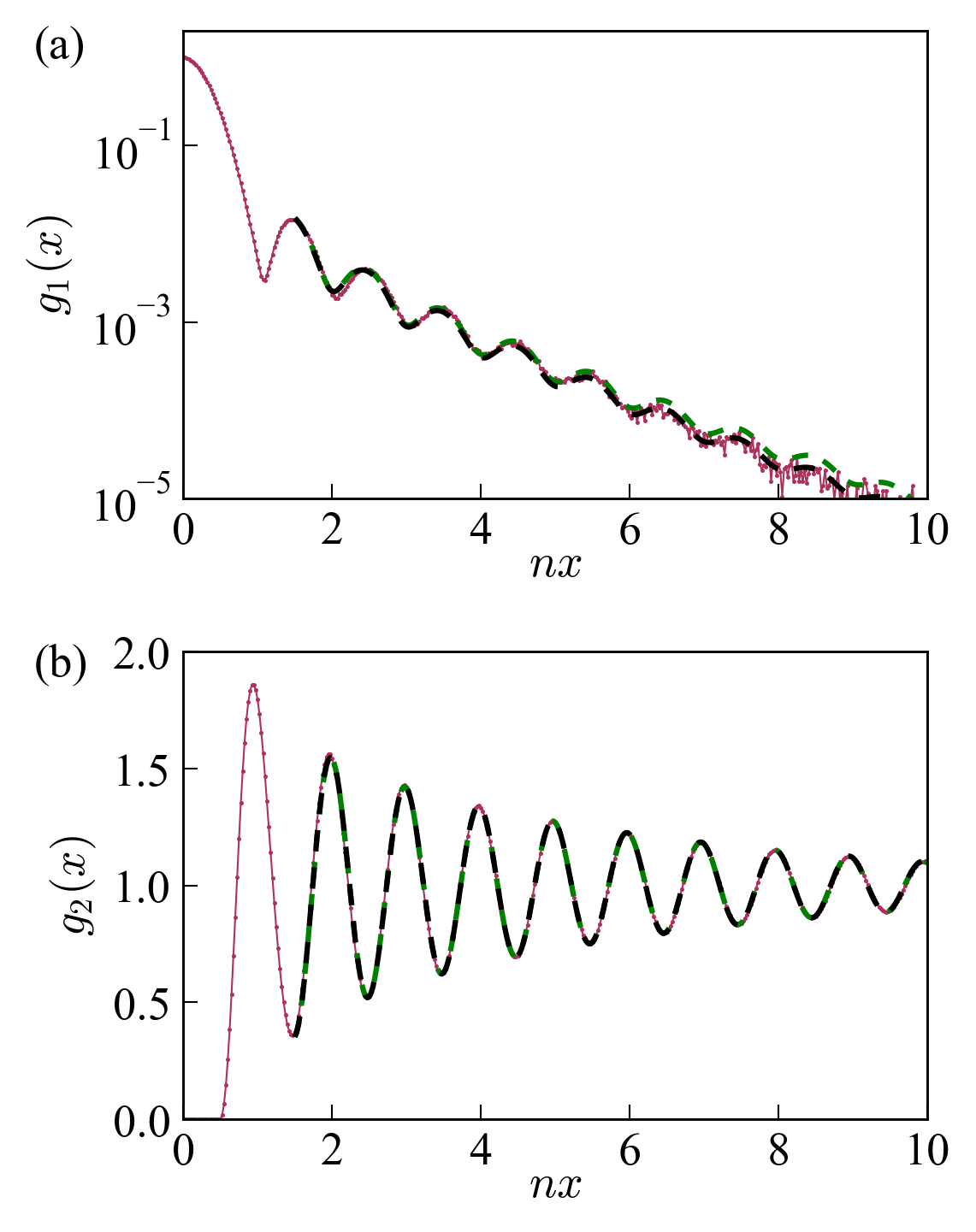}
\caption{Fits of correlation functions. The solid lines are $g_1(x)$ in panel~(a) and $g_2(x)$ in panel~(b) for $n\aHR=0.5$ and $T=0.4\hbar^2/ma^2\kB$, while the dashed lines are the fitted curves at step one (dashed green line) and step two (dashed black line).}
\label{fig:fits_appendix}
\end{figure}

\renewcommand{\theequation}{C.\arabic{equation}}
Here we show a typical result of the fits of correlation functions.
Figure~\ref{fig:fits_appendix} shows the one-body correlation function $g_1(x)$ and the pair correlation function $g_2(x)$ for $n\aHR=0.5$ and $T=0.4\hbar^2/ma^2\kB$ as found by QMC calculations (red dots and lines), together with the corresponding fits (dashed green line for the first-step fit and dashed black line for the second-step fit).
We find similarly good agreement for all fits performed for this work.
\FloatBarrier

\bibliographystyle{revtexlsp}
\bibliography{bibYSJ, notes}

\end{document}